\documentclass[aps,prl,twocolumn,reprint,showpacs,superscriptaddress,footinbib,groupedaddress]{revtex4-1}
\usepackage{graphicx} 
\usepackage[francais]{babel}
\usepackage{dsfont} 
\usepackage{amsmath} 
\usepackage{verbatim} 
\usepackage{graphicx} 
\usepackage[utf8]{inputenc} 
\usepackage{ulem} %added to be able to cross text out 
\usepackage{units} % for using nicefrac in texts 
\usepackage{version} % for comment 
\usepackage{color} 
\usepackage{colortbl} 
\usepackage{xcolor}  % for writting parts in color, e.g. in red 
\usepackage{fancyref} 
\usepackage[colorlinks=true,citecolor=black,urlcolor=blue,linkcolor=black]{hyperref} % choice of color for hyperref links
\usepackage{array} 
\usepackage{booktabs} 
\usepackage{longtable}

\begin{document} 

\title{Stability analysis of a magnetic waveguide with self-generated offset field}  
\author{C. L. Garrido Alzar} 
\email{carlos.garrido@obspm.fr} 

\affiliation{SYRTE, Observatoire de Paris, Universit\'e PSL, CNRS, Sorbonne Universit\'e, LNE, 61 avenue de l'Observatoire, 75014 Paris, France}
\date{\today} 

\begin{abstract} 
The unexpected emerging stability of a time-modulated magnetic guide, realized without external offset fields, is 
demonstrated. We found a steady periodic solution around which the nonlinear dynamics is linearized. To investigate 
the orbital stability of the guiding, a formal criterion based on the analysis of the eigenvalues of the monodromy matrix 
of the system dynamics is used. To circumvent the difficulty in finding an analytical expression for these eigenvalues, a 
Lyapunov transformation of the system variables is proposed. From this transformation an equation of state for the 
system parameters is derived, and it allows to estimate an upper bound for the computed stability domains. In particular, 
we found a general expression for the threshold modulation frequency below which the guiding is unstable. Using 
experimentally accessible parameters, the stable guiding of $^{87}$Rb atoms is investigated.
\end{abstract} 

\maketitle

%%%%%%%%%%%%%%% %%%% Guide design %%%%%%%%%%%%%%%%%%%%%%%%%%%%%%%%%%%%%%%%%%%%%%%%%%%%%%%%%%%%
The actual technological progress allows the miniaturization of sensors to a level where the 
laws of the quantum world control the working principles of these devices. Among the outstanding 
realizations we can cite the SQUID magnetometers~\cite{squid}, the electrical resistance standard based on the 
quantum Hall effect~\cite{qhe}, and the superconducting gravimeter~\cite{igrav}. In the last 
decades important efforts have been carried out for the application of matter wave interferometers which 
belong to another class of quantum devices using cold neutral atoms~\cite{muquans,aosense,gillot,canuel}. In this sense, when 
the development of compact matter wave interferometers 
is considered, atom chips come up as a promising technology for the manipulation of cold atoms using complicated
geometries~\cite{keil}. Indeed, it is possible to microfabricate on an atom chip a complex wire pattern to create miniature 
magnetic potentials with the shape required by the targeted application. We can, for example, design arrays of 
potential wells for quantum information processing~\cite{sinuco,wang,leung}, traps for acceleration 
measurements~\cite{ammar}, and toroidal waveguides for rotation sensing~\cite{clga,lesanovsky,fernholz,seungjin}.

So far all the classical realizations of magnetic potentials, in free space or on atom chips, use a homogeneous offset 
magnetic field to lift the degeneracy between the confined and not confined atomic states. Generally, a pair of 
macroscopic coils in a Helmholtz configuration is used to achieve this goal. However, this method is neither practical 
for the development of a compact device nor compatible with miniature magnetic potentials of complex shapes.

In particular, for a ring guiding potential of a rotation sensor, the use of coils is the simplest and straightforward 
way to obtain the offset field that fits the symmetry of the potential. However, the switching time of the potential 
in this situation is undesirably increased limiting the device operation. In fact, the use of coils runs against the 
realization of high bandwidth and low power consumption sensors, two key performance ingredients for embedded applications 
such as inertial navigation. This extra timing, added to the already important time required to prepare the samples between 
each interferometric measurement, will degrade the stability via the Dick effect~\cite{dick}.

Here, a solution to generate the offset field of a magnetic toroidal waveguide that takes into account the 
above mentioned problems is proposed. The considered magnetic waveguide is generated by three concentric microwires 
fabricated on an atom chip. The currents in the inner $I_2$ and outer $I_3$ microwires will be assumed to flow in the same 
direction, and opposite to the current $I_1$ in the central microwire. This currents will be modulated at a frequency 
$\omega$, as in the configuration used to suppress the magnetic potential roughness (see Fig. 1 in~\cite{trebbia}). However, 
instead of setting and trying to keep an exact phase difference of $\pi$ between the currents $I_2=I_3$ and $I_1$, 
here an small phase offset $\phi \ll \pi$ is introduced so that the total phase difference is 
$\pi+\phi:$  $I_1(t) = A_1 \cos (\omega t)$ and $I_2(t) = I_3(t)=A_2 \cos (\omega t +\pi +\phi)$.

The existence of $\phi$ has two main consequences on the magnetic guiding potential. The first one is the presence of a 
residual roughness~\cite{trebbia,bouchoule} that could affect the propagation of atoms close to the surface. This question 
is beyond the scope of the present work and will be analyzed in detail elsewhere. Nevertheless, this 
is not a limiting factor for the proposed solution because we can always find a minimum working distance from the atom chip 
surface where the effects of the roughness on the propagation are totally negligible (see, for example, Fig. 63 
in~\cite{fortagh}). This distance is typically on the order of a few hundreds of microns in the atom chip experiments 
dealing with matter wave interferometry. The proposed guide with self-generated offset field is {\it also} compatible 
with these working distances. The second consequence is that the obtained time-varying magnetic potential has a 
dynamical minimum with zero field that goes through 
the initial atom cloud location, and this fact raises the question about the non-intuitive and unexpected emerging 
stability of this waveguide against atom losses, to be demonstrated below.

\begin{figure}[htb]
 \begin{center}
 \includegraphics[width=0.35\textwidth]{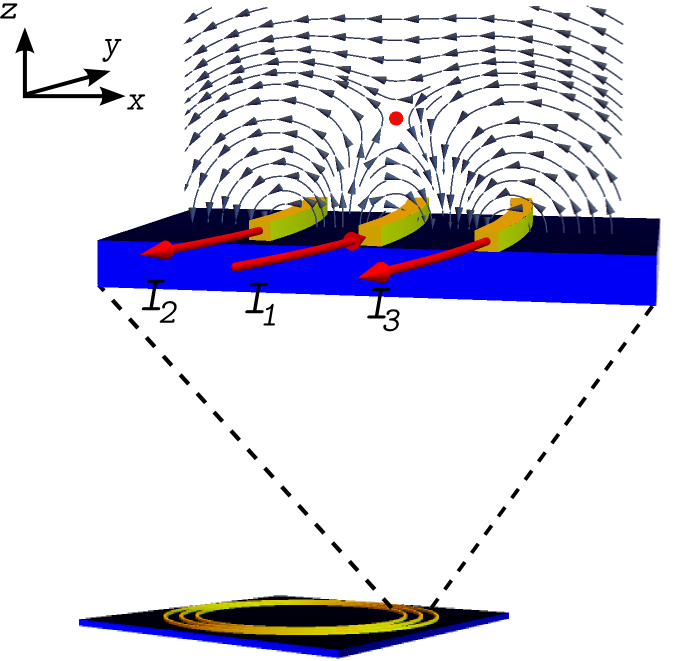}
 \caption{(color online). Current configuration generating the magnetic guide. In the zoom, the considered linear section of 
 the circular guide is represented together with the magnetic field lines map created by the currents. The red point in this map 
 indicates the guide minimum at a given time instant.}
 \label{fig:3w}
 \end{center}
\end{figure}

In this paper, an attempt to approach the analytical answer to this stability question is presented. To simplify the 
mathematical treatment without loss of generality, we will consider a linear section of a ring magnetic guide with a radius 
much larger than the microwires separation $l$, as shown in Fig.~\ref{fig:3w}. Let us assume that the currents flow in 
the $y$ direction of a local reference frame, centered at the field minimum (red point in the field lines map above the 
central wire), for $\phi=0$ and $t=0$, in the $z$ direction perpendicular to the chip surface. 
Then the dynamics of the atoms in the $x-z$ plane is described by the equations~\cite{gov1,gov,franzosi}
\begin{align}
 M\frac{d^2 x}{dt^2} & =\mu \frac{\partial}{\partial x} [\mathbf{n}\cdot\mathbf{B}(\mathbf{r},t)]\label{eq:dyn1}\ ,\\
 M\frac{d^2 z}{dt^2} & =\mu  \frac{\partial}{\partial z} [\mathbf{n}\cdot\mathbf{B}(\mathbf{r},t)]\label{eq:dyn2}\ ,\\
 \hbar\frac{d\mathbf{n}}{dt}&=\mu \mathbf{n}\times \mathbf{B}(\mathbf{r},t)\label{eq:dyn3}\ ,
\end{align}
where $M$, $\mu$ and $\mathbf{n}$ are the mass of an atom, its magnetic moment and a unit vector in the direction of the 
magnetic moment, respectively. The magnetic field produced by the microwires $\mathbf{B}(\mathbf{r},t)$ can be written to 
first order in the position coordinates, and for $\phi \ll \pi$, as~\cite{fortagh} 
\begin{equation}
 \mathbf{B}(\mathbf{r},t)=[b z\cos(\omega t)+\phi B_{\rm b}\sin(\omega t)]\mathbf{i} + b x\cos(\omega t)\mathbf{k}\ ,
 \label{eq:btmod}
\end{equation} 
where $B_{\rm b}$ is the magnetic field produced by the inner and outer microwires only, and $b$ is the field 
gradient. Introducing
the dimensionless positions and velocities $X \equiv x/l$, $Z \equiv z/l$, $\tau \equiv \omega t$, $V_x \equiv v_x/l\omega$, 
and $V_z \equiv v_z /l\omega$, the time evolution of the external and internal degrees of freedom of an atom in 
the magnetic guide with self-generated offset field can be described by the expressions
\begin{align}
\ddot{X}&=-\alpha_1 n_z \cos(\tau)\nonumber\ ,\\
\ddot{Z}&=-\alpha_1 n_x \cos(\tau)\nonumber\ ,\\
\dot{n}_x&=-\alpha_2 n_y  X\cos(\tau)\ ,\label{eq:NLDE}\\
\dot{n}_y&=-\alpha_2 [n_z Z-n_x X]\cos(\tau)- \alpha_3 n_z\sin(\tau)\nonumber\ ,\\
\dot{n}_z&=\alpha_2 n_y Z\cos(\tau)+ \alpha_3 n_y\sin(\tau)\nonumber\ ,
\end{align}
where for simplicity we have omitted the time dependence of the system's variables. In these equations 
${\bf n(\tau)}=(n_x,n_y,n_z)$ will give the instantaneous direction of the atom's magnetic moment and the 
dimensionless constants $\alpha_i$ are defined by the equations
\begin{align}
 \alpha_1 \equiv \frac{g_F \mu_B b}{M \omega^2 l} &= \frac{\omega_\perp^2}{\omega^2}\ ,\ 
 \alpha_2 \equiv \frac{g_F \mu_B b l}{\hbar \omega} = \frac{\Omega}{\omega}\nonumber\ ,\\
 \alpha_3 \equiv &\frac{g_F \mu_B \phi B_{\rm b}}{\hbar \omega} = \frac{\omega_L}{\omega}\ ,
\label{eq:as}
\end{align}
where $g_F$ and $\mu_B$ are respectively the Land\'e factor and the Bohr's magneton. In the quantum mechanical situation,
we need to consider Eqs.~(\ref{eq:as}) describing the atoms with the total angular momentum $\mathbf{F}$ and magnetic 
moment $\mu=-g_F \mu_B \mathbf{F}/\hbar$.

Together with the definitions of the $\alpha_i$ parameters we also introduced 
the three characteristic frequencies that determine the different time scales involved in this dynamics: the Larmor 
frequency of the atoms in the field produced by the two external microwires, 
$\omega_L \equiv g_F \mu_B \phi B_{\rm b}/\hbar$; the characteristic frequency of the atomic transverse motion, 
$\omega_\perp \equiv \sqrt{g_F \mu_B b/M l}$; and the characteristic Rabi frequency at which the magnetic moment 
couples to the modulated field gradient, $\Omega \equiv g_F \mu_B b l/\hbar$. Using these characteristic frequencies, the 
solutions expressed in terms of the $\alpha$'s can be applied to any magnetically trappable atomic species.

The Eqs.~(\ref{eq:NLDE}) represent a system of 
nonlinear differential equations (NDEs). As usually, we will start the analysis of its stability by 
linearizing it around a steady periodic solution or orbit. Indeed, it turns out that some properties of the linearized system 
are also valid for the original nonlinear system, in particular the stability. Thus, we shall seek 
for periodic solutions in the form
\begin{align}
 X(\tau) &\equiv X_c(\tau)\cos(\tau)+X_s(\tau)\sin(\tau)\ ,\label{eq:X}\\
 Z(\tau) &\equiv Z_c(\tau)\cos(\tau)+Z_s(\tau)\sin(\tau)\ ,\label{eq:Z}
\end{align}
based on the fact that we have harmonic driving of the system's variables. In Eqs.~(\ref{eq:X}) and~(\ref{eq:Z}) 
the envelopes $X_c(\tau), X_s(\tau), Z_c(\tau), Z_s(\tau)$ are supposed to be slowly varying when compared to 
$\sin(\tau)$ and $\cos(\tau)$. In addition, since by definition at any time we must have ${\bf n}^2(\tau)=1$, we will 
assume the following form for the components of this unit vector
\begin{align}
 n_x(\tau) &=\cos[\theta(\tau)]\sin[\nu(\tau)]\ ,\label{eq:nx}\\
 n_y(\tau) &=\sin[\theta(\tau)]\sin[\nu(\tau)]\ ,\label{eq:ny}\\
 n_z(\tau) &=\cos[\nu(\tau)]\ .\label{eq:nz}
\end{align}

Substituting Eqs.~(\ref{eq:X})-(\ref{eq:nz}) in Eqs.~(\ref{eq:NLDE}), neglecting high order derivatives of the slowly 
varying envelopes, and equating the coefficients in front of the $\sin(\tau)$ and $\cos(\tau)$ terms in the 
RHS and LHS of the position equations, we arrive finally at the following system of first order NDEs
\begin{align}
\dot{X_c}&=-\frac{1}{2}X_s \nonumber\ ,\\
\dot{X_s}&=\frac{1}{2}X_c - \frac{\alpha_1}{2} \cos(\nu) \nonumber\ ,\\
\dot{Z_c}&=-\frac{1}{2}Z_s \nonumber\ ,\\
\dot{Z_s}&=\frac{1}{2}Z_c-\frac{\alpha_1}{2} \cos(\theta)\sin(\nu) \ ,\label{eq:NLDE2}\\
\dot{\theta}&=-\cos(\theta)\cot(\nu)\{\alpha_2[Z_c\cos^2(\tau)+Z_s\sin(\tau)\cos(\tau)]\nonumber\\
      &+\alpha_3\sin(\tau)\}+\alpha_2[X_c\cos^2(\tau)+X_s\sin(\tau)\cos(\tau)]\nonumber\ ,\\
\dot{\nu}&=-\sin(\theta)\{\alpha_2[Z_c\cos^2(\tau)+Z_s\sin(\tau)\cos(\tau)]\nonumber\\
      &+\alpha_3\sin(\tau)\}\ .\nonumber
\end{align}

It is no difficult to find the steady periodic solutions of equations~(\ref{eq:NLDE2}) which are 
given by $\theta_k=k\pi$, $\nu_m=(2m+1)\pi/2$, $X_c = X_s = Z_s = 0$, and $Z_c=(-1)^{k+m}\alpha_1$ with $k$ and $m$ 
integers. If we linearize Eqs.~(\ref{eq:NLDE2}) around this steady orbit, then we obtain the following matrix of time 
varying coefficients to describe the linearized dynamics
\begin{widetext}
\begin{equation}
 A(\tau)=\left( \begin{array}{cccccc}
 0 & -1/2 & 0 & 0 & 0 & 0 \\
 1/2 & 0 & 0 & 0 & 0 & (-1)^m \alpha_1/2 \\
 0 & 0 & 0 & -1/2 & 0 & 0 \\
 0 & 0 & 1/2 & 0 & 0 & 0 \\
 \alpha_2 \cos^2(\tau) & \alpha_2 \sin(\tau)\cos(\tau) & 0 & 0 & 0 &  f(\alpha_1,\alpha_2,\alpha_3,\tau) \\
 0 & 0 & 0 & 0 & - f(\alpha_1,\alpha_2,\alpha_3,\tau) & 0
 \end{array} \right) \ ,
\label{eq:Amatrix}
\end{equation}
\end{widetext}
with
\begin{equation}
 f(\alpha_1,\alpha_2,\alpha_3,\tau) \equiv (-1)^k [(-1)^{k+m}\alpha_1 \alpha_2\cos^2(\tau)+\alpha_3 \sin(\tau)]\ .
\label{eq:f}
\end{equation}

Now, let us introduce the formal mathematical framework for the analysis of the qualitative behavior of the linearized system.
Because of the harmonic terms appearing in Eqs.~(\ref{eq:NLDE}) and~(\ref{eq:Amatrix}), and because of the periodic 
solutions we are interested in, here we are concerned with the orbital stability as defined by Theorem~7.4 
in~\cite{verhulst}. In addition, the matrix $A(\tau)$ is continuous $2\pi$-periodic and consequently, by 
virtue of the Floquet theorem the fundamental matrix $\Phi(\tau)$ of $A(\tau)$ can be written as a 
product of two matrices and it satisfies the equation~\cite{lukes}
\begin{equation}
 \Phi(\tau+2\pi)=\Phi(\tau)\mathcal{M}\ .
\label{eq:fms}
\end{equation}
The matrix $\mathcal{M}$ is called the monodromy matrix (or Floquet multipliers matrix) and its eigenvalues determine the 
stability behavior of~(\ref{eq:NLDE2}). Indeed, a necessary and sufficient condition for orbital stability of the periodic 
solution is that all eigenvalues of $\mathcal{M}$ have modulus smaller than 1. This is the criterion that will be used 
to investigate the stability by linearization of our system of NDEs.

If $\Phi(\tau)$ is a fundamental matrix solution with $\Phi(0)=I$, being $I$ the identity matrix, then we can find 
$\mathcal{M}$ and 
its eigenvalues from Eq.~(\ref{eq:fms}). Unfortunately, the matrix $A(\tau)$ is not commutative on $[0,+\infty)$ and 
hence it is 
not possible to easily find a closed form for $\Phi(\tau)$ that can give some insights on the physics and the underlying 
structure of the periodic solution. Still, an approximated solution can be found using a P\'eano-Baker series as given 
by the Lemma~2.1 in~\cite{peano}, namely
\begin{eqnarray}
 \Phi(\tau)=I+\int^\tau_0 dt_1 A(t_1)+\int^\tau_0 dt_1 A(t_1) \int^{t_1}_0 dt_2 A(t_2)\nonumber\\
  +\int^\tau_0 dt_1 A(t_1) \int^{t_1}_0 dt_2 A(t_2)\int^{t_2}_0 dt_3 A(t_3)+\cdots\ .
\label{eq:fmss}
\end{eqnarray}

The above presented formalism is summarized in the following algorithm used to check the stability of 
our magnetic waveguide with self-generated offset field:\\
{\it Step 1}: Find the steady periodic solution of Eqs.~(\ref{eq:NLDE2}) and linearize this system around this orbit.\\
{\it Step 2}: Using the matrix $A(\tau)$ resulting from {\it Step 1}, compute the fundamental matrix solution from 
Eq.~(\ref{eq:fmss}).\\
{\it Step 3}: For a given point in the parameter space $\{\alpha_1,\alpha_2,\alpha_3\}$, compute the monodromy matrix 
$\mathcal{M}$ using~(\ref{eq:fms}).\\
{\it Step 4}: Find the eigenvalues of $\mathcal{M}$ for the parameter values chosen in {\it Step 3}. If {\it all} these 
eigenvalues are smaller than 1 in modulus, then the system is orbitally stable in this point; otherwise, it is not.

\begin{figure}[htb]
 \begin{center}
 \includegraphics[width=0.47\textwidth]{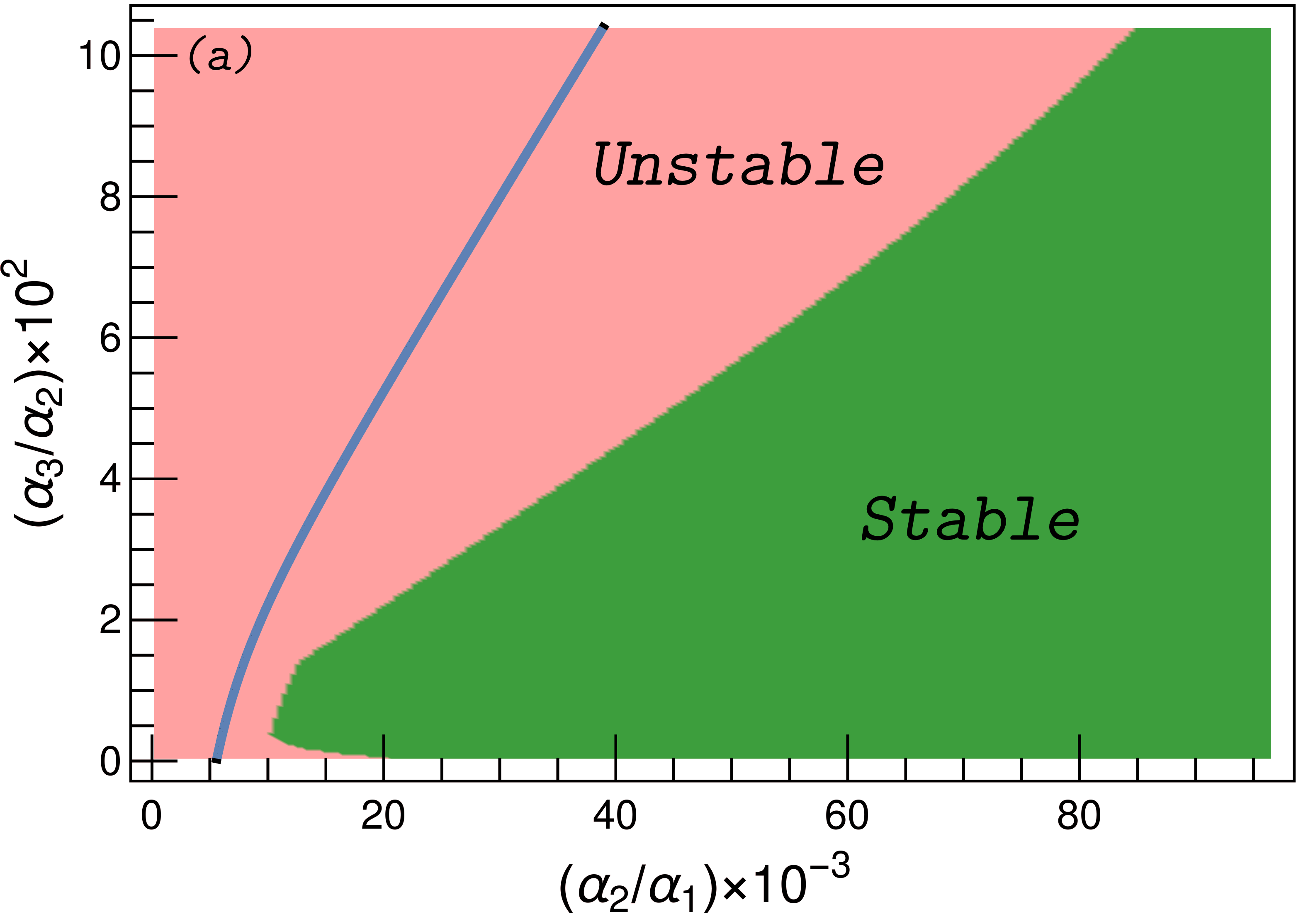}\vskip 0mm
 \includegraphics[width=0.47\textwidth]{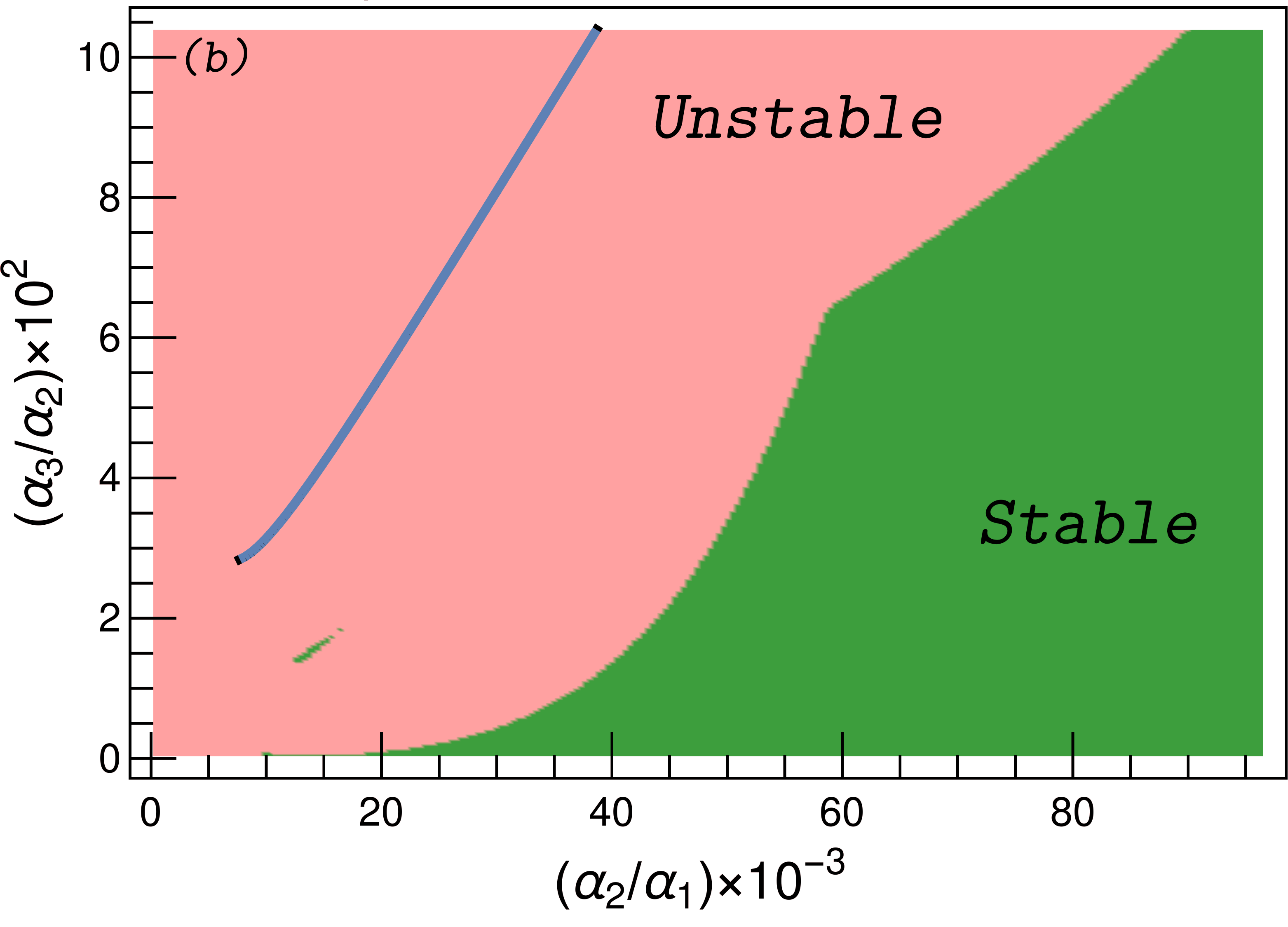}
 \caption{(color online). Stability phase diagram of the magnetic waveguide with self-generated 
 offset field, for (a) $k=1$, $m=0$ and (b) $k=m=1$. The blue line is an estimation of the upper bound 
 of the stability domain computed with Eq.~(\ref{eq:bound}). The threshold frequency corresponds to 
 $\alpha_2/\alpha_1=\sqrt[3]{\Omega^4/2\omega_\perp^4}\approx 6\times10^3$.}
 \label{fig:phdiag}
 \end{center}
\end{figure}

Notice that this algorithm is based only on the analysis of the structure of NDEs and thus, it is general and can be 
applied to any matrix $A(\tau)$. As an example, let us investigate the guiding of 
$^{87}$Rb atoms, trapped initially in the $|F=1, m_F=2\rangle$ state, in a magnetic guide with a self-generated offset field 
produced by three microwires separated by 
$l=15\ \mu$m, with $B_{\rm b} \approx$ 1.5 G and $b \approx$ 290 G/cm. The stability phase diagram calculated 
with these realistic experimental parameters is shown in Fig.~\ref{fig:phdiag}. This figure is obtained using a fourth 
order P\'eano-Baker 
series and 0.1\% precision. We found 
that for $m$ even, the phase diagrams with $k$ even, $\alpha_3<0$ and 
$k$ odd, $\alpha_3>0$ coincide. By inspecting the matrix $A(\tau)$ we see that this is an expected result. 
In addition, the steady periodic solution indicates that the magnetic moment of the atom is aligned along the 
$x$ axis since $\theta_k=k\pi$ and $\nu_m=(2m+1)\pi/2$. In fact this is not surprising, after all the self-generated 
offset field $\phi B_{\rm b}$ is oriented along this axis.
The phase diagram shown in Fig.~\ref{fig:phdiag} represents the global stability of the investigated system. This implies 
that the actual extent of the stable domain, for a given situation, will be defined by how far the initial conditions are 
from the steady periodic solution around which we linearize the system dynamics. In other words, the atoms need to be 
injected in the stable orbits sustained by the guide.

The calculation of the stability boundary requires the knowledge of the analytical expressions of the eigenvalues 
of $\mathcal{M}$, which are very difficult to calculate for the actual problem. Therefore, the physical mechanisms 
participating in 
the stable guiding of the atoms are not easily accessible. Nevertheless, we can gain some intuition by applying the 
Lyapunov transformation $L(\tau)$ 
\begin{equation}
 L(\tau)=\left( \begin{array}{cccccc}
 1 & 0 & 0 & 0 & 0 & 0 \\
 0 & 1 & 0 & 0 & 0 & 0 \\
 0 & 0 & 1 & 0 & 0 & 0 \\
 0 & 0 & 0 & 1 & 0 & 0 \\
 0 & 0 & 0 & 0 & \cos[\beta(\tau)] & -\sin[\beta(\tau)] \\
 0 & 0 & 0 & 0 & \sin[\beta(\tau)] & \cos[\beta(\tau)]
 \end{array} \right) \ ,
\label{eq:Lmatrix}
\end{equation}
to the Eq.~(\ref{eq:Amatrix}), with $\beta(\tau)$ defined by the equation
\begin{equation}
 \dot{\beta}(\tau)+f(\alpha_1,\alpha_2,\alpha_3,\tau) =1\ .
\label{eq:beta}
\end{equation}

The Lyapunov transformation~(\ref{eq:Lmatrix}) does not reduce the linearized system given by Eq.~(\ref{eq:Amatrix}) 
to a system with constant coefficients as it is usually desired~\cite{kirillov}. However, here is shown that it is enough for 
this transformation to be of Lyapunov-type to obtain the physical behavior of the system. Indeed, a Lyapunov transformation
does not change the characteristic exponents of a linear system and preserve its regularity.

The function $\beta(\tau)$ is constrained to be periodic satisfying the condition $\beta(\pi)=\beta(2\pi)$, that translates 
into the following equation of state for the parameters of the system
\begin{equation} 
 \pi=(-1)^k \Big[\frac{(-1)^{k+m}}{2}\pi\alpha_1 \alpha_2-2\alpha_3\Big]\ .
\label{eq:bound}
\end{equation}

\begin{figure}[htb]
 \begin{center}
 \includegraphics[width=0.47\textwidth]{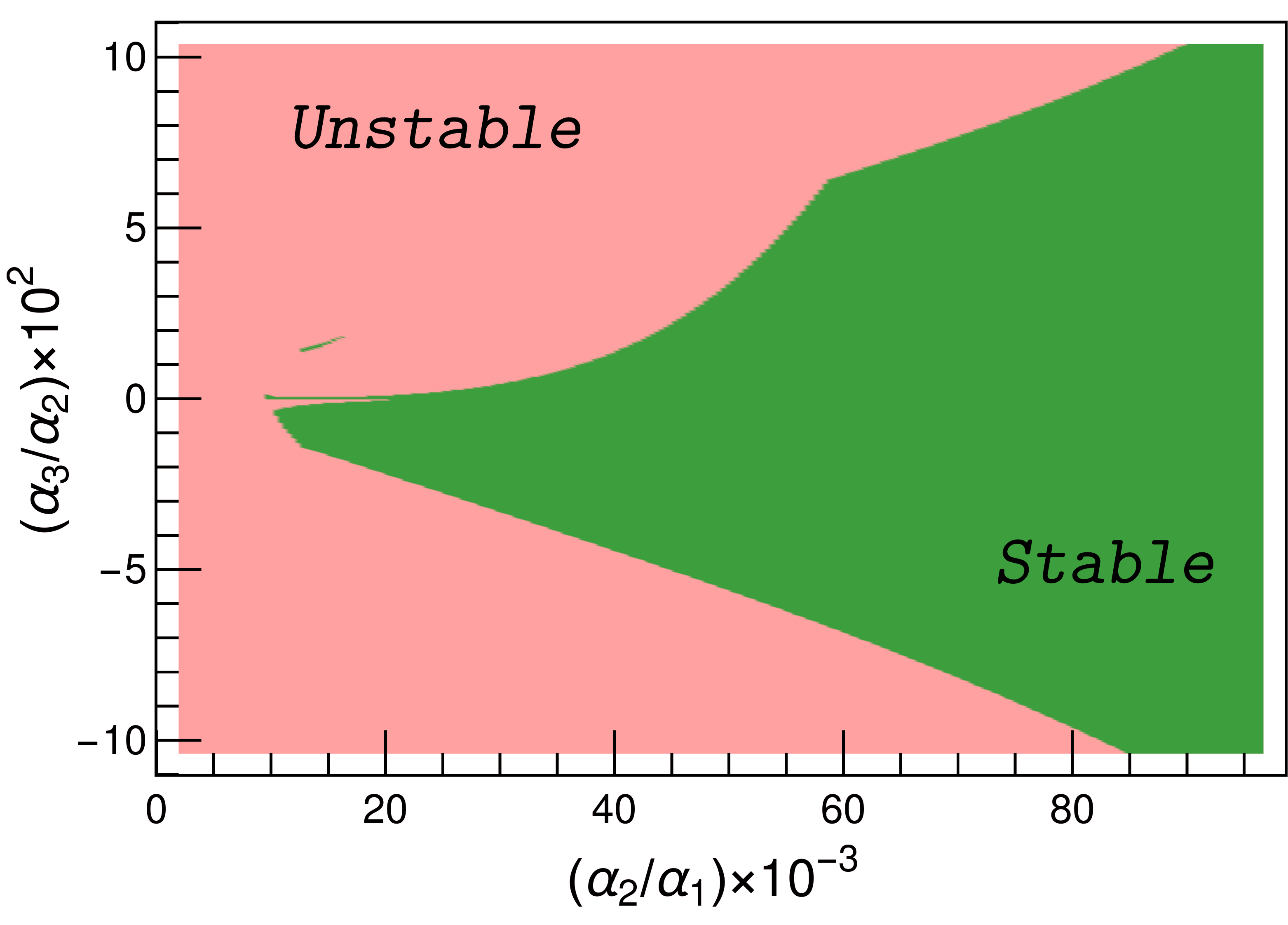}
 \caption{(color online). Stability phase diagram for $k=m=0$.}
 \label{fig:phdiag1}
 \end{center}
\end{figure}

Then, using Eqs.~(\ref{eq:as}), we can obtain the valid pairs of values ($\omega,\omega_L)$ which set an upper bound for the 
stability domain for the case $k=1$, $m=0$ considered, for example, in Fig.~\ref{fig:phdiag}(a). Indeed, in this particular 
case the Eq.~(\ref{eq:bound}) reads
\begin{equation} 
 \omega^2(\omega-\frac{2}{\pi}\omega_L) - \frac{1}{2}\omega_\perp^2 \Omega=0\ ,
\label{eq:bound1}
\end{equation}
from where we can conclude that since $\omega_\perp^2 \Omega$ is always positive we must have 
$\omega > (2/\pi)\omega_L$. In Fig.~\ref{fig:phdiag}(a) we have 
represented the curve~(\ref{eq:bound1}) by the solid line 
which starts at $\omega_{th}=\sqrt[3]{\omega_\perp^2 \Omega/2}$, approximately equal to $2\pi\times 3$ kHz for the 
parameter values considered in this particular example. Below 
this threshold frequency, the dynamics is unstable independently of the value of $\phi$. The same conclusion can be drawn from 
Fig.~\ref{fig:phdiag}(b) calculated for $k=m=1$, where the same threshold frequency is observed. Therefore, the stable 
guiding in these situations 
requires a modulation frequency above the Larmor frequency. Finally, in 
Fig.~\ref{fig:phdiag1} the stability phase diagram covering positive and negative values of $\alpha_3$ (or $\phi$) 
is presented for $k=0$, $m=0$. Comparing this figure with Fig.~\ref{fig:phdiag}, we can notice the expected 
symmetry between $k$ even, $\alpha_3<0$ and $k$ odd, $\alpha_3>0$.

Here, we have demonstrated using formal arguments the existence of guiding stability domains in a magnetic waveguide 
created by a linear section of three concentric microwires carrying modulated currents. The striking point of this waveguide 
is that, in the 
absence of an external offset field, it can be made orbitally stable against atom losses by properly choosing the modulation 
frequency and the phase relationship between the currents. This unexpected emerging stability is 
established by analyzing the eigenvalues of the monodromy matrix of the dynamics linearized around a steady periodic 
solution. We found that the stability results from the interplay between 
the characteristic frequencies, as seen from the estimated limiting conditions obtained from a Lyapunov transformation 
of the system variables. More specifically, we found a lower bound for the modulation frequency determined by the Larmor 
frequency of the atoms. The next important question to be investigated is the coherence of matter waves during propagation 
in such a guide. We believe this work to be of relevance not only when considering the design of toroidal magnetic 
guides for high bandwidth rotation sensing with atom chips, but also for the design of complex miniature 
fast switching magnetic potentials for the study of low dimensional physics using atom chip 
based devices~\cite{clga,crookston,fernholz}.

%%%%%%%%%%%%%%%%%%%%%%%%%%%%%%%%%%%%%%%%%%%%%%%%%%%%%%%%%%%%%%%%%%%%%%%%%%%%%%
This work was funded by the D\'el\'egation G\'en\'erale de l'Armement (DGA)
through the ANR ASTRID program (contrat ANR-13-ASTR-0031-01), the Institut Francilien de Recherche 
sur les Atomes Froids (IFRAF), the Emergence-UPMC program (contrat A1-MC-JC-2011/220).

%%%%%%%%%%%%%%%%%%%%%%%%%%%%%%%%%%%%%%%%%%%%%%%%%%%%%%%%%%%%%%%%%%%%%%%%%%%%%%%

\end{document}